\documentclass[slac_one]{revtex4}
\usepackage{graphicx}
\usepackage{fancyhdr}
\begin{document}

\title{\boldmath{$b$}-Tagging Algorithms and their Performance at ATLAS} 

\author{M.~Lehmacher (on behalf of the ATLAS collaboration)}
\affiliation{Physikalisches Institut, University of Bonn, Nussallee 12, 53115 Bonn, Germany}

\begin{abstract}
The ability to identify jets containing $B$ hadrons is important for the high-$p_T$ physics program of a general-purpose experiment such as ATLAS. $b$-tagging is in particular useful for selecting very pure top quark samples, for studying standard model or supersymmetric Higgs bosons which couple preferably to heavy objects, for vetoing backgrounds for several physics channels and finally for searching for new physics beyond the standard model like supersymmetric particles or heavy gauge bosons. After a review of the algorithms used to identify $b$-jets, their anticipated performance is discussed as well as the impact of selected critical ingredients such as residual misalignments in the tracker. The prospects to measure the $b$-tagging performance in the first few hundreds pb$^{-1}$ of data with di-jet events as well as $t\overline{t}$ events are presented. Finally three different physics use cases of $b$-tagging are summarised.
\end{abstract}

\maketitle

\thispagestyle{fancy}

\section{\boldmath{$b$}-TAGGING ALGORITHMS AT ATLAS} 

Bottom jets possess several characteristic properties that can be utilised for separating them from jets coming from the hadronisation of lighter quarks. The most important property is the relatively long lifetime of $B$ hadrons of the order of $1.5\,$ps, which leads to a measurable flight length path of typically a few millimeters before the subsequent decay into lighter hadrons. A significant fraction of $B$ hadrons decay semi-leptonically into hadrons containing charm quarks, which then decay further into lighter hadrons. The decay of the $B$ hadrons at a displaced secondary vertex can be identified inclusively by measuring the impact parameters of tracks coming from the decay, that is the distance from the point of closest approach of the track to the interaction vertex. The impact parameter itself is a signed quantity, where it is chosen positive if the point of closest approach lies upstream with respect to the jet direction and negative in the other case. Apart from that, the secondary vertex can be reconstructed explicitly, either inclusively or exclusively, the latter implying a full reconstruction of the $B$ hadron decay chain. In addition the relatively large mass of more than $5\,$GeV produces  opening angles and transverse momenta of the decay products relative to the $B$ hadron flight direction that differ significantly from other jets. The various tagging methods studied at ATLAS can be divided into two main classes. The \emph{spatial taggers} encompass methods that utilise lifetime information like impact parameters and decay vertices. The \emph{soft-lepton taggers} on the other hand are based on the reconstruction of the lepton in case the $B$ hadron decayed semi-leptonically. These non-isolated leptons have a sizable tranverse momentum as well as a large transverse momentum relative to the jet axis $p_T^{rel}$. The first property comes from the fact that the $b$ quark fragmentation is \emph{hard}, that is the $B$ hadron retains a major fraction of the original $b$ quark momentum. Detailed information on the presented results and $b$-tagging at ATLAS in general can be found in the chapter on $b$-tagging in \cite{CSC} and references therein.

\begin{figure}[htbp]
\centering
\begin{minipage}[b]{7 cm}
\includegraphics[width=60mm]{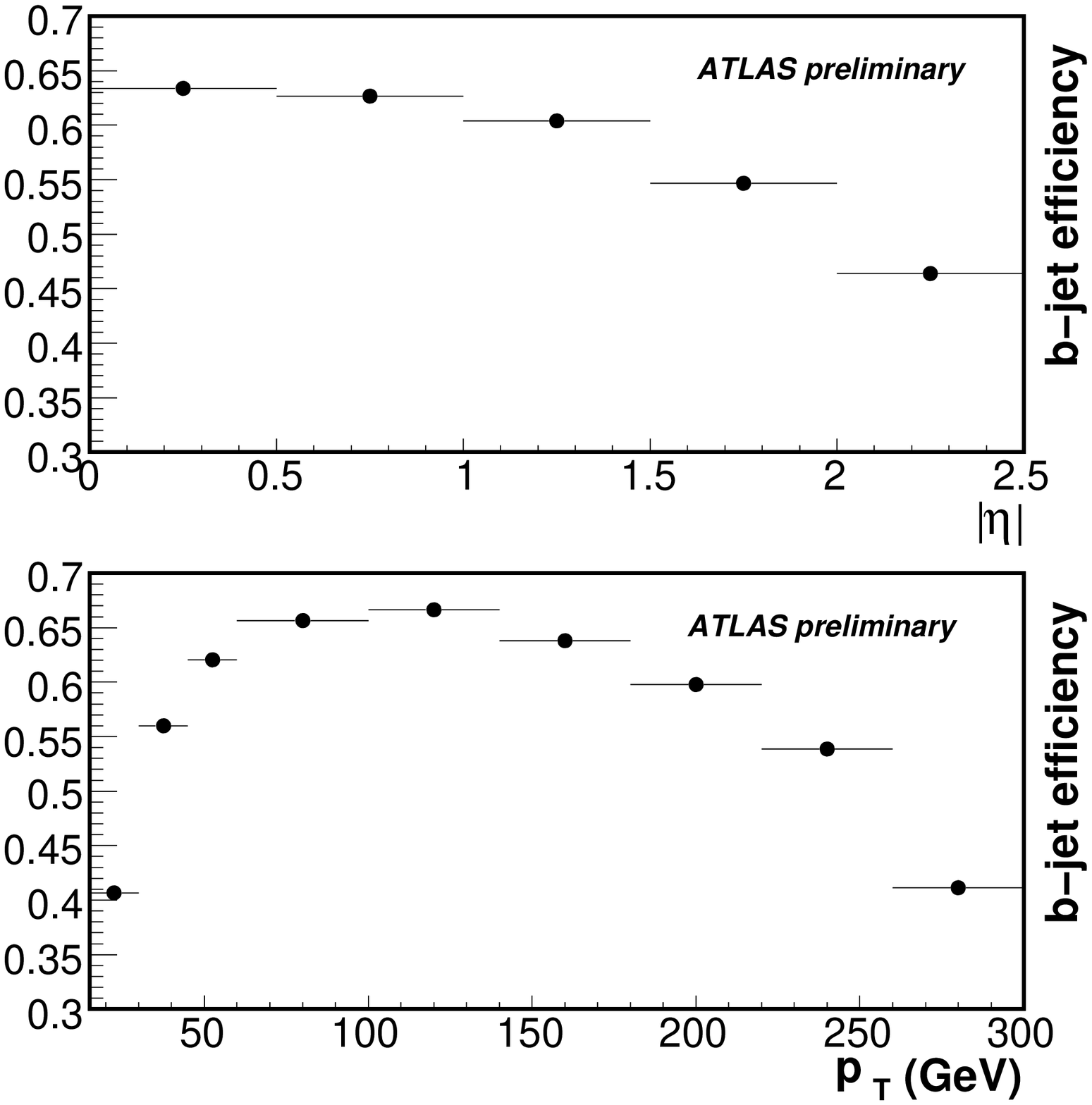}
\end{minipage}
\begin{minipage}[b]{8 cm}
\includegraphics[width=60mm]{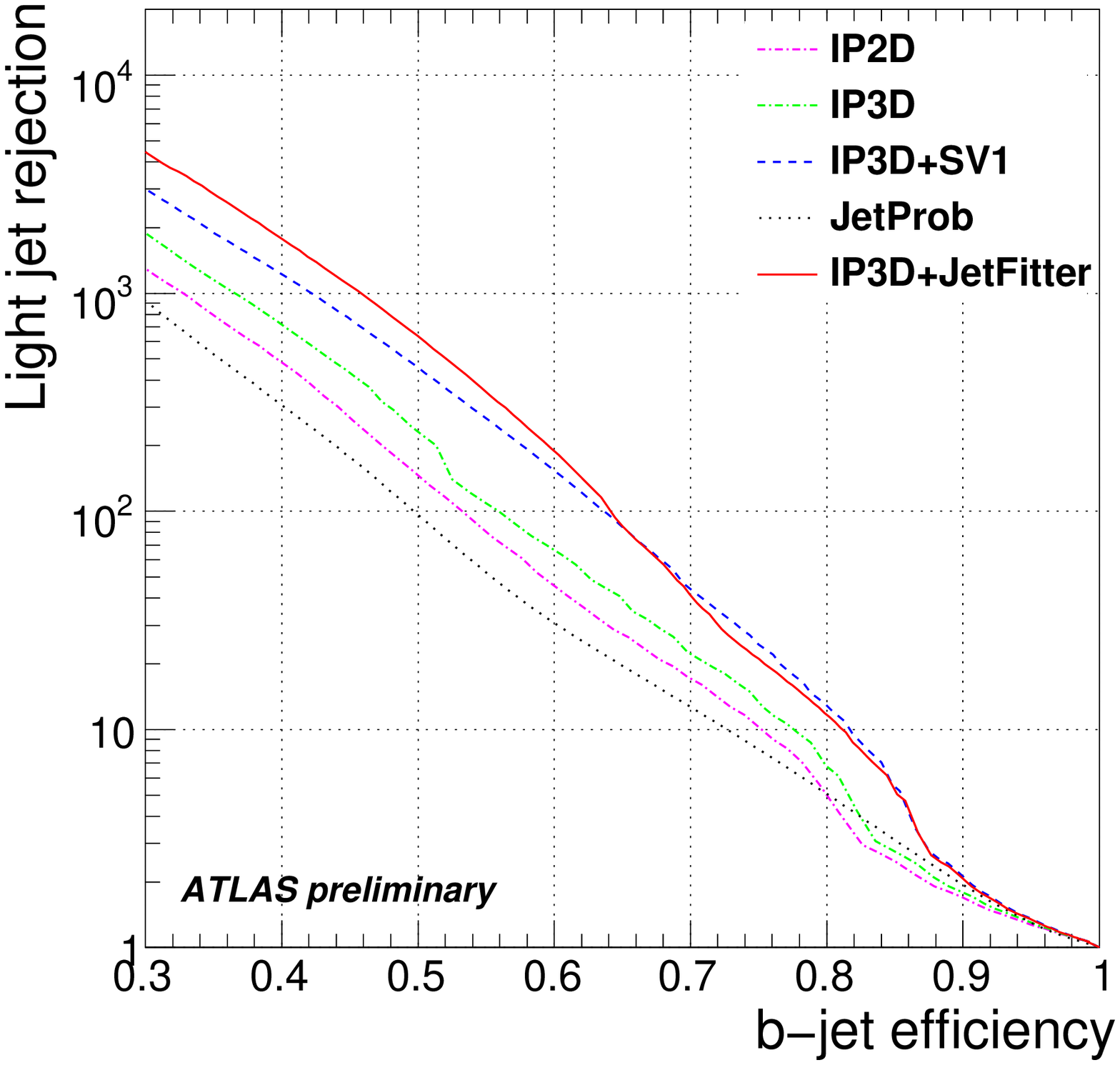}
\end{minipage}
\caption{Left: $b$-tagging efficiency obtained with the IP3D+SV1 tagging algorithm operating at a fixed cut on the $b$-tagging weight for $t\overline{t}$ events versus jet $p_T$ and jet $|\eta|$ respectively\cite{CSC}. Right: Rejection of light jets versus $b$-jet efficiency for $t\overline{t}$ events and for the following tagging algorithms: JetProb, IP2D, IP3D, IP3D+SV1, IP3D+JetFitter\cite{CSC}.} \label{perf}
\end{figure}

Apart from a few simple taggers, that are intended mainly for the early start-up phase of the experiment where one prefers simple, robust and easy to calibrate taggers, all tagging methods rely on the likelihood ratio approach to build a discriminating variable, called jet weight, for the separation of $b$-jets, $c$-jets and other jets. For this purpose one or more dimensional reference histograms of discriminating variables are constructed with a Monte Carlo simulation for the different hypotheses. In some cases the weights of different taggers are combined to reach better discrimination. In the following only the separation between $b$-jets and light-jets is considered for simplicity. All jets having a jet weight above a certain cut value are then tagged as $b$-jets. The specific cut value determines the $b$-tagging efficiency $\epsilon_b$, which is defined as the fraction of true $b$-jets that are tagged as $b$-jets, as well as the rejection of light-jets $R_u$. The latter is defined as the inverse of the fraction of true light-jets that are falsely tagged as $b$-jets. For a given cut on the weight the rejection of light jets as well as the efficiency, in general, strongly depend on $\eta$ and $p_T$ of the jet, see Figure \ref{perf}.

The simplest taggers just tag a jet as a $b$-jet, if either there are at least $N$ tracks associated to the jet with impact parameter (IP) larger than a fixed value (\emph{TrackCounting}), or the $B$ hadron flight distance significance as calculated from the secondary vertex position exceeds some cut value (\emph{SV0}). Another tagger (\emph{JetProb}) uses the negative side of the (transverse) IP significance distribution as obtained from prompt tracks to calculate the probability of compatibility of the tracks in the jet with the primary vertex. This probability is then used as jet weight. As this tagger uses only information from prompt tracks it is expected to be simpler to calibrate than the more sophisticated likelihood ratio taggers, which in addition use reference distributions for $b$-jets. A more sophisticated impact parameter based tagger, that uses likelihood ratios, called \emph{IPxD}, utilises the distributions of the IP significance as calculated in the longitudinal projection (\emph{IP1D}), the transverse projection (\emph{IP2D}) or a combination of both in two-dimensional histograms (\emph{IP3D}). Both \emph{IPxD} and \emph{JetProb} divide all tracks into categories with dedicated reference distributions before calculating the jet weight to improve the performance.

One secondary-vertex tagger (\emph{SV1/2}) fits inclusive secondary vertices and builds the jet weight from several one- or more-dimensional variable distributions like vertex mass, energy fraction of the tracks fitted to the vertex to all tracks in the jet, number of tracks in the vertex and angle between jet direction and $B$ hadron flight direction as estimated from the primary-secondary vertex axis. In addition, the resulting jet weight is combined with the \emph{IP3D} tagger. The tagger with the best performance (\emph{JetFitter}) fits the decay chain of $B$ hadrons, i.e. it fits a common $B$/$C$ flight direction along with the position of additional vertices on it. This even includes \emph{incomplete} topologies where no vertex could be fitted. The jet weight is calculated similarly to \emph{SV1/2}, but taking different decay topologies into account via categories. The \emph{JetFitter} algorithm can also be combined with \emph{IP3D} for optimal performance.

Two soft-lepton tagger approaches are pursued at ATLAS. One uses soft muons and one or two dimensional reference histograms of the muonic $p_T$ and the muonic $p_T^{rel}$. The other uses electrons and relies on the challenging identification of soft electrons inside jets.

\section{PERFORMANCE IN MONTE CARLO SIMULATIONS}

The performance of the tagging methods is estimated on Monte Carlo simulated events. For the following results $t\overline{t}$ events were used, where jets were reconstructed via a seeded cone algorithm based on calorimeter towers with a jet cone radius of $\Delta R=0.4$. A true $b$-jet is then defined as any reconstructed jet for which there exists at least one $b$ quark with $p_T>5\,$GeV in a cone of $\Delta R=0.3$ around the jet axis. Light jets are jets where neither a $b$ quark nor $c$ quarks nor tau leptons could be matched to the jet in this way. Furthermore, all jets have to pass the preselection criteria $p_T>15\,$GeV and $|\eta|<2.5$. Tracks are associated to jets also via a $\Delta R$ criterion with a default value of $\Delta R=0.4$. A snapshot of the expected light jet rejection as a function of the efficiency can be found for the spatial taggers in Figure \ref{perf}. One can expect $R_u\sim 30$ for the tagger \emph{JetProb} up to $R_u\sim 200$ for the sophisticated \emph{JetFitter} at a typical $b$-tagging efficiency of $\epsilon_b=60\%$. The soft muon tagger  for example gives $R_u\sim 300$ at $\epsilon_b=10\%$, where $\epsilon_b$ includes branching ratios. A separation of charm jets and $b$-jets is much more difficult due to very similar properties of both types of jets leading to a rejection that is typically a factor of $\sim10$ smaller than for light jets. Since the \emph{JetFitter} algorithm deals with the actual topology of the decay, it handles $c$-jets better than the other methods reaching a charm rejection of $\sim20$. It is interesting to note that $b$-tagging is possible on trigger level with a light jet rejection of $\sim20$ at $60\%$ at the last trigger stage.

An effort has been made to reach a realistic understanding of critical aspects of $b$-tagging. The studies summarised above where done assuming a perfect knowledge of all misalignments. More realistic studies take residual misalignments into account as well as the process of realignment including systematic errors. Recent studies indicate a possible degradation of the light jet rejection of up to $\sim25\%$ at most for fixed $\epsilon_b$, but not all systematic effects have been taken into account yet. Further degradations in rejection are seen in studies including pile-up events ($\sim5$ minimum bias events are expected at an instantaneous luminosity of $2\cdot 10^{33}\,$cm$^{-2}$s$^{-1}$), where in a few percent of the cases a wrong primary vertex is reconstructed leading, besides other things, to an artificial shift in the longitudinal IP and finally to a loss in rejection of $\sim20-30\%$ for \emph{IP3D} and \emph{IP3D+SV1}. Studies are carried out to recover the lost performance by optimising the $b$-tagging for the case of pile-up. The taggers that take only the transverse impact parameter into account are not affected by pile-up though. 

\section{PROSPECTS FOR PERFORMANCE MEASUREMENTS}

It will be necessary to calibrate the $b$-tagging methods on data. This means that one has to measure the tagging efficiency as well as the mis-tagging efficiency, i.e. the inverse of the rejection. Due to the dependence on $p_T$ and $\eta$ it is desirable to perform the calibration in bins of those variables. In addition one would also like to measure the used reference histograms with data. For the efficiency measurement several methods have been studied at ATLAS that make use of either di-jet or $t\overline{t}$ events.

The \emph{$p_T^{rel}$} method uses events with jets that include non-isolated muons. Templates of the muon $p_T^{rel}$ as obtained from simulated and reconstructed $b$-, $c$- and light-jets passing basic selection criteria are fitted to the measured distribution before and after applying the respective tagger, which preferably is a spatial tagger. By counting the number of muonic jets before and after the tagging, the efficiency $\epsilon_b$ can be estimated. Another method that uses muonic jets is called \emph{System 8}. This method needs two event samples with different flavor composition, preferably enriched with $b$-jets, as well as two uncorrelated tagging methods. For the first sample one selects jets with non-isolated muons, whereas the other is a subsample with one additional b-tagged jet in the opposite hemisphere. For the taggers one can use a spatial one and a soft lepton tagger. By solving eight equations with eight unknowns one is able to determine the $b$-tagging efficiency. A special trigger for muonic jets was implemented to collect the necessary amount of events and by using different trigger thresholds and prescaling it is possible to use these methods in bins of $p_T$ and $\eta$ with a fixed amount of data. It was shown that for both methods an integrated luminosity of $\sim50\,$pb$^{-1}$ is sufficient to derive detailed calibration curves in $p_T$ or $\eta$. The relative precision of the absolute value of $\epsilon_b$ is estimated for both methods to be at the $6\%$ level, which is limited by systematic errors.  

There are several methods that make use of $t\overline{t}$ events. First, there is an event counting method, that measures the average $b$-tagging efficiency and the cross section of $t\overline{t}$ production in the lepton+jets or the dilepton channel at the same time, by counting the number of events with 1, 2 or 3 tagged jets. With $\sim100\,$pb$^{-1}$ of data a relative precision of $\sim(2.2$(stat.)$\pm3.5$(sys.)$)\%$ can be reached in the lepton+jets channel. In order to measure the efficiency in bins of $p_T$ and $\eta$ one identifies $b$-jets by fully reconstructing the $t\overline{t}$ decay chain in the lepton+jets channel. The classification is improved by requiring a b-tag for one jet coming from the hadronically decaying top. The $b$-tagging is then studied on the other unbiased side. Three methods are explored for selecting such samples, a kinematic fit, a likelihood selection and a topological selection. With $\sim200\,$pb$^{-1}$ of data a relative precision of $\sim(7.7$(stat.)$\pm3.2$(sys.)$)\%$ can be reached for example for the topological method. With a larger amount of data the latter techniques could also be used to extract the reference distributions for $b$-jets.

\section{PHYSICS USE CASES}

There are various examples of use cases where $b$-tagging is a critical ingredient. One is the search for a standard model Higgs boson in the $t\overline{t}H (H\rightarrow b\overline{b})$ channel. Here $b$-tagging can be used to reduce or even eliminate large background like $t\overline{t}j\overline{j}$ or $W+$jets. For example the $t\overline{t}j\overline{j}$ background is reduced by two orders of magnitude by using $b$-tagging. Using $b$-tagging the signal significance is estimated to be $S/\sqrt{B}\sim 2.2$ at an integrated luminosity of $\sim30\,$fb$^{-1}$. Another example is the top mass measurement. There the highest precision can be reached in the $t\overline{t}$ lepton+jets channel by requiring two b-tagged jets using the hadronically decaying top as the mass estimator. Assuming a jet energy scale uncertainty of the order of one percent, a precision of $1\,$GeV can be reached with an integrated luminosity of $1\,$fb$^{-1}$, which then would be already dominated by the systematic errors. A complementary approach relying on $b$-tagging infers the top mass from the mean transverse decay length of $B$ hadrons coming from the top decays. Here the uncertainty due to the jet energy scale is negligible. Finally $b$-tagging is also important for exotic models, where one has to reconstruct very high-$p_T$ jets above $500\,$GeV. These jets are challenging for tracking as well as $b$-tagging, due to the high track density and multiplicity as well as the fact, that the $B$ hadron decay lengths are so large that the secondary vertex can be located around or after the first detector layers.

\section{CONCLUSIONS}

Various algorithms for tagging $b$-jets have been studied in detail at ATLAS. The spectrum covers simple, robust taggers as well as sophisticated taggers, that make use of as much information as possible from the $B$ hadron decay chain. Several approaches for calibrating $b$-tagging algorithms with data were shown to be realisable with a few $100\,$pb$^{-1}$ of data. $b$-tagging is essential for many physics analyses, like Higgs boson searches or top mass measurements.

\begin{acknowledgments}
The author wishes to thank M. Cristinziani, L. Vacavant and N. Wermes.
\end{acknowledgments}

\end{document}